\documentclass[aps,prb,reprint,superscriptaddress,amsmath,amssymb]{revtex4-2}
\usepackage{dcolumn}
\usepackage{bm,graphicx}
\usepackage{etoolbox}
\usepackage{url}
\usepackage{booktabs}
\usepackage[colorlinks]{hyperref}
\usepackage{breakurl}
\usepackage{color}
\usepackage[dvipsnames]{xcolor}
\usepackage [utf8]{inputenc}
\usepackage[T1]{fontenc}
\usepackage{siunitx}
\usepackage{comment}
\DeclareSIUnit{\angstrom}{\textup{\AA}}
\begin{document}

\title{Room-temperature antiferromagnetic resonance in NaMnAs}

\author{Jan Dzian}
\email{jan.dzian@lncmi.cnrs.fr}
\affiliation{Faculty of Mathematics and Physics, Charles University, Prague}
\affiliation{Laboratoire National des Champs Magn\'etiques Intenses, LNCMI-EMFL, CNRS UPR3228, Univ.~Grenoble Alpes, Univ.~Toulouse, Univ.~Toulouse~3, INSA-T, Grenoble and Toulouse, France}

\author{St\'a\v{n}a~T\'azlar\r{u}}
\affiliation{Faculty of Mathematics and Physics, Charles University, Prague}
\affiliation{Institute of Physics, Academy of Science of the Czech Republic, Cukrovarnick\'a 10, Praha 6, CZ-16253, Czech Republic}

\author{Ivan Mohelsk\'y}
\affiliation{Laboratoire National des Champs Magn\'etiques Intenses, LNCMI-EMFL, CNRS UPR3228, Univ.~Grenoble Alpes, Univ.~Toulouse, Univ.~Toulouse~3, INSA-T, Grenoble and Toulouse, France}
\affiliation{Department of Physics, University
of Fribourg, CH-1700 Fribourg, Switzerland}

\author{Florian Le~Mardel\'e}
\affiliation{Laboratoire National des Champs Magn\'etiques Intenses, LNCMI-EMFL, CNRS UPR3228, Univ.~Grenoble Alpes, Univ.~Toulouse, Univ.~Toulouse~3, INSA-T, Grenoble and Toulouse, France}

\author{Filip~Chudoba}
\affiliation{Laboratoire National des Champs Magn\'etiques Intenses, LNCMI-EMFL, CNRS UPR3228, Univ.~Grenoble Alpes, Univ.~Toulouse, Univ.~Toulouse~3, INSA-T, Grenoble and Toulouse, France}

\author{Ji\v{r}\'i Voln\'y}
\affiliation{Faculty of Mathematics and Physics, Charles University, Prague}

\author{Jan Wyzula}
\affiliation{Laboratoire National des Champs Magn\'etiques Intenses, LNCMI-EMFL, CNRS UPR3228, Univ.~Grenoble Alpes, Univ.~Toulouse, Univ.~Toulouse~3, INSA-T, Grenoble and Toulouse, France}
\affiliation{Department of Physics, University
of Fribourg, CH-1700 Fribourg, Switzerland}

\author{Amit Pawbake}
\affiliation{Laboratoire National des Champs Magn\'etiques Intenses, LNCMI-EMFL, CNRS UPR3228, Univ.~Grenoble Alpes, Univ.~Toulouse, Univ.~Toulouse~3, INSA-T, Grenoble and Toulouse, France}

\author{Simone~Ritarossi}
\affiliation{Sapienza Universit\'a di Roma, Department of Physics, Roma, Italy}
\affiliation{Dipartimento di Scienze, Università degli Studi di Roma Tre, Via della Vasca Navale 84, 00146 Roma, Italy}
\affiliation{INFN Sezione di Roma Tre, Via della Vasca Navale 84, 00146 Roma, Italy}

\author{Riccardo~Mazzarello}
\affiliation{Sapienza Universit\'a di Roma, Department of Physics, Roma, Italy}

\author{Philipp Ritzinger}
\affiliation{Institute of Physics, Academy of Science of the Czech Republic, Cukrovarnick\'a 10, Praha 6, CZ-16253, Czech Republic}

\author{Jakub~\v Zelezn\'y}
\affiliation{Institute of Physics, Academy of Science of the Czech Republic, Cukrovarnick\'a 10, Praha 6, CZ-16253, Czech Republic}

\author{Karel V\'yborn\'y}
\affiliation{Institute of Physics, Academy of Science of the Czech Republic, Cukrovarnick\'a 10, Praha 6, CZ-16253, Czech Republic}

\author{Kl\'ara Uhl\'i\v{r}ov\'a}
\affiliation{Faculty of Mathematics and Physics, Charles University, Prague}

\author{Beno\^it Gr\'emaud}
\affiliation{Aix Marseille Univ, Universit\'e de Toulon, CNRS, CPT, Marseille, France}

\author{Andr\'es Sa\'ul}
\affiliation{Aix-Marseille Universit\'e, Centre Interdisciplinaire de Nanoscience de Marseille-CNRS (UMR 7325), Marseille, France}

\author{Cl\'ement Faugeras}
\affiliation{Laboratoire National des Champs Magn\'etiques Intenses, LNCMI-EMFL, CNRS UPR3228, Univ.~Grenoble Alpes, Univ.~Toulouse, Univ.~Toulouse~3, INSA-T, Grenoble and Toulouse, France}

\author{Martin Veis}
\email{martin.veis@matfyz.cuni.cz}
\affiliation{Faculty of Mathematics and Physics, Charles University, Prague}

\author{Milan Orlita}
\email{milan.orlita@lncmi.cnrs.fr}
\affiliation{Laboratoire National des Champs Magn\'etiques Intenses, LNCMI-EMFL, CNRS UPR3228, Univ.~Grenoble Alpes, Univ.~Toulouse, Univ.~Toulouse~3, INSA-T, Grenoble and Toulouse, France}
\affiliation{Faculty of Mathematics and Physics, Charles University, Prague}

\date{\today}

\begin{abstract}

We report on antiferromagnetic resonance experiments in bulk tetragonal NaMnAs –- a room-temperature antiferromagnetic semiconductor. Our results corroborate previous ab initio studies, which propose that NaMnAs is an easy-axis antiferromagnet with the N\'eel vector oriented along the tetragonal axis. At $B=0$, we find a single antiferromagnetic resonance line at 7~meV and associate it with a doubly degenerate ($k = 0$) magnon mode. Its energy softens considerably with increasing $T$, but remains clearly visible in the data up to room temperature. From the experimental data, we estimate the single-ion anisotropy of the Mn ions in NaMnAs in the range 0.1-0.2~meV, a value that is relatively large compared to other manganese-based antiferromagnets.
\end{abstract}


\maketitle

\section{Introduction}
\label{intro}

Studies of magnetically ordered van der Waals (vdW) systems constitute a significant and rapidly growing area of research in solid-state physics, driven both by fundamental interest and potential applications. However, the vast majority of investigated systems exhibit magnetic ordering only at low temperatures.
Prominent examples include the MX\textsubscript{3} family (M = Cr, Ru, V; X = Cl, Br, I) \cite{Kuhlow1982-oy,Tsubokawa1960-xo,Dillon1965-lu,Fletcher1967,Juza1969-br} and the MPX\textsubscript{3} family (M = Mn, Fe, Co, Ni; X = S, Se) \cite{Okuda1986-yl,Taylor1974-bm,Brec1986-xz}. Consequently, there remains a strong ongoing search for layered materials that exhibit magnetic order at room temperature.

\begin{figure}[ht!]
    \centering
    \includegraphics[width=0.7\linewidth]{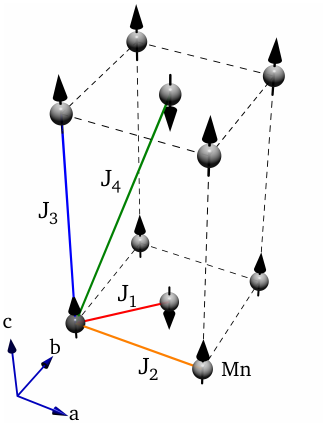}
    \caption{Schematic spin structure of NaMnAs magnetic crystal cell with only the magnetic Mn\textsuperscript{2+} ions visible. The spins of Mn\textsuperscript{2+} are aligned parallel to the tetragonal axis (c axis). The colored lines represent nearest neighbor exchange interactions of different orders. J\textsubscript{n} indicates the nearest neighbor couplings, with $n$ being the order of the coupling.}
    \label{fig:namnas-structure}
\end{figure}

The relatively limited body of knowledge accumulated so far on NaMnAs~\cite{Bronger1986,VolnyPRB22}, indicates that this material is an antiferromagnetic semiconductor, with ordering temperature $T_N$ around 350~K. This contrasts with half-Heusler (cubic) NaMnAs expected theoretically to be a ferromagnetic spin-gapless semiconductor~\cite{HaRSCA25}. Tetragonal NaMnAs is a layered compound that can be easily cleaved along the crystallographic plane perpendicular to the tetragonal axis~\cite{VolnyPRB22}. The mechanical properties of NaMnAs thus resemble, to a great extent, those of well-known vdW systems. However, the preparation of thin flakes\textemdash down to a few atomic layers or even a monolayer\textemdash has not yet been experimentally demonstrated. According to ab initio calculations~\cite{ZhouJMMM16,VolnyPRB22}, NaMnAs is predicted to be a $C$-type antiferromagnet~\cite{Blundel01}, characterized by ferromagnetic chains running along the tetragonal axis (see Fig.~\ref{fig:namnas-structure}). Similar to genuine van der Waals materials, the exchange coupling between adjacent layers is expected to be significantly weaker than the in-plane exchange interaction~\cite{VolnyPRB22}.


In this paper, we investigate low-energy excitations in bulk NaMnAs using frequency-domain THz spectroscopy under applied magnetic fields. At low temperatures, we identify a doubly-degenerated magnon mode in the THz spectral range with behavior typical of an easy-axis antiferromagnet. With increasing temperature, this excitation progressively softens, but remains clearly detectable, even at room temperature.

\section{Experimental details}
\label{Exp}

The NaMnAs single crystals studied were grown using the flux growth technique, described in detail in~Ref.~\cite{VolnyPRB22}, and prior to experiments, were kept in an evacuated quartz ampoule to avoid surface oxidation. Nevertheless, since solely bulk properties are targeted in our optical experiments, no relevant differences were found in the observed response of surface-degraded or freshly cleaved samples. The NaMnAs crystals, as-grown, are relatively soft and have plate-like shapes, with the surface parallel to cleavable planes, \emph{i.e.}, perpendicular to the tetragonal ($c$) axis. For our optical experiments, we used as-grown samples, with no additional wedging. 

\begin{figure}[t]
    \centering
    \includegraphics[width=0.9\linewidth]{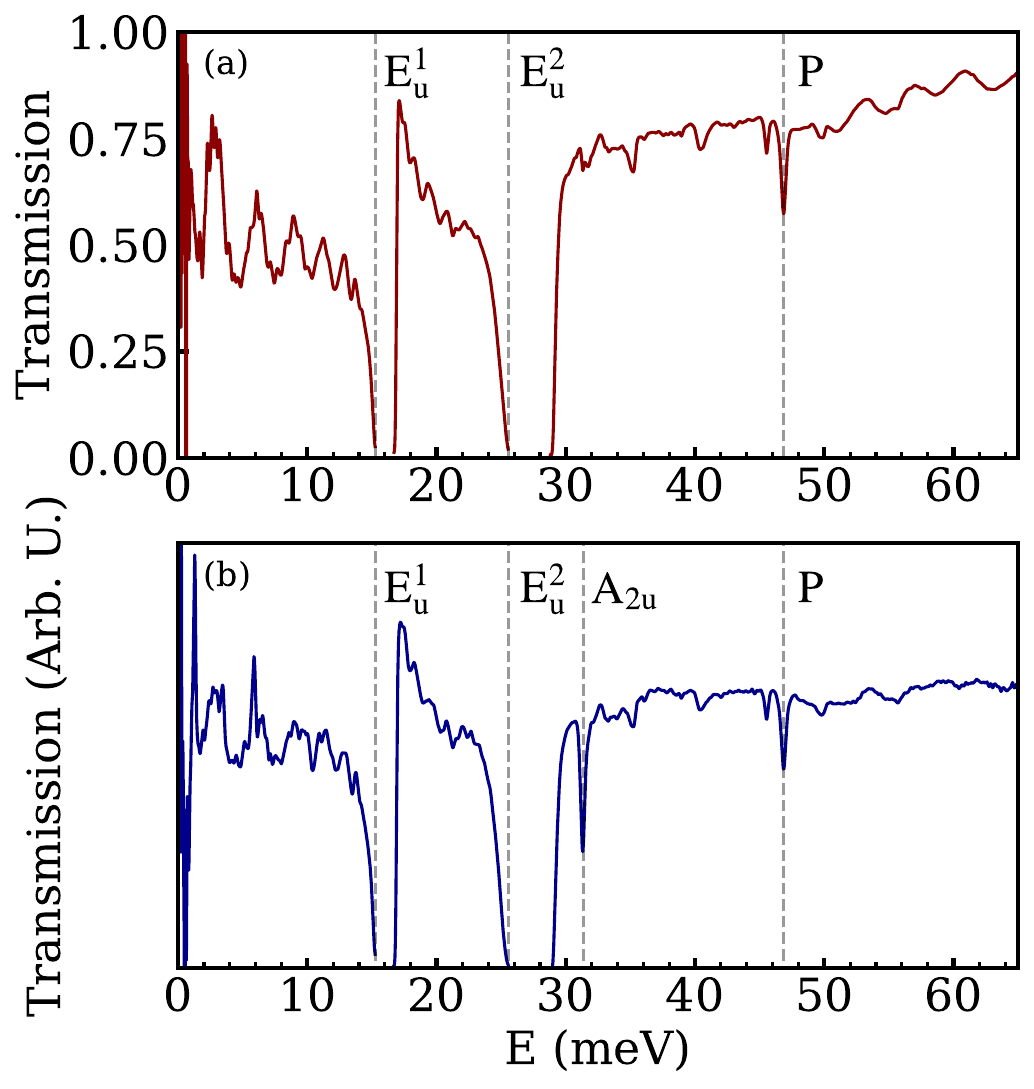}
    \caption{Low-temperature (T=4.2~K) transmission spectrum of a bulk NaMnAs crystal, several hundred microns thick, in the THz-infrared spectral range. The wave vector is aligned with the tetragonal axis in (a) and tilted by 35$^{\circ}$ in (b). Vertical dashed lines show positions of IR active phonons.}
    \label{fig:namnas_art_irphonons}
\end{figure}

The THz-infrared transmission experiments, the results of which are presented below in Figs.~\ref{fig:namnas_art_irphonons} ,\ref{fig:namnas_IR_faraday}, \ref{fig:namnas_IR_voigt} and \ref{fig:namnas_IR_temp}, were realized using Vertex 80v Fourier transform spectrometer. The radiation from a mercury lamp was guided using light-pipe optics to the sample placed inside the superconducting or resistive coils (up to 16 and 30~T, respectively) and surrounded by the helium exchange gas. Three distinct configurations have been employed: (i) magneto-transmission in the Faraday geometry, with $B$ applied along the tetragonal axis of NaMnAs, see Fig.~\ref{fig:namnas_IR_faraday}. In this configuration, the sample was mounted on a rotation stage that enabled referencing by an open aperture. This allowed us to get also absolute transmission in Fig.~\ref{fig:namnas_art_irphonons}; (ii) magneto-transmission in the Voigt geometry, with $B$ applied perpendicular to the tetragonal axis, see Fig.~\ref{fig:namnas_IR_voigt}; (iii) magneto-transmission in the Faraday geometry, with the NaMnAs sample placed on a gold mirror, for measurement at a temperature varying in the range 4.2-295~K, see Fig.~\ref{fig:namnas_IR_temp}. 

In the arrangement (iii), the beam reflected by the mirror, passing effectively twice through the sample at varying temperature, was detected by an external bolometer. In configurations (i) and (ii), the signal was collected using a bolometer placed just below the sample, both kept at the base temperature of $T=4.2$~K. Throughout the paper, we use the notation $B_\|$ and $B_\perp$ that indicates the direction of the applied magnetic field along or perpendicular to the tetragonal axis of NaMnAs ($c$-axis). 

\begin{figure*}[ht!]
    \centering
    \includegraphics[width=0.98\linewidth]{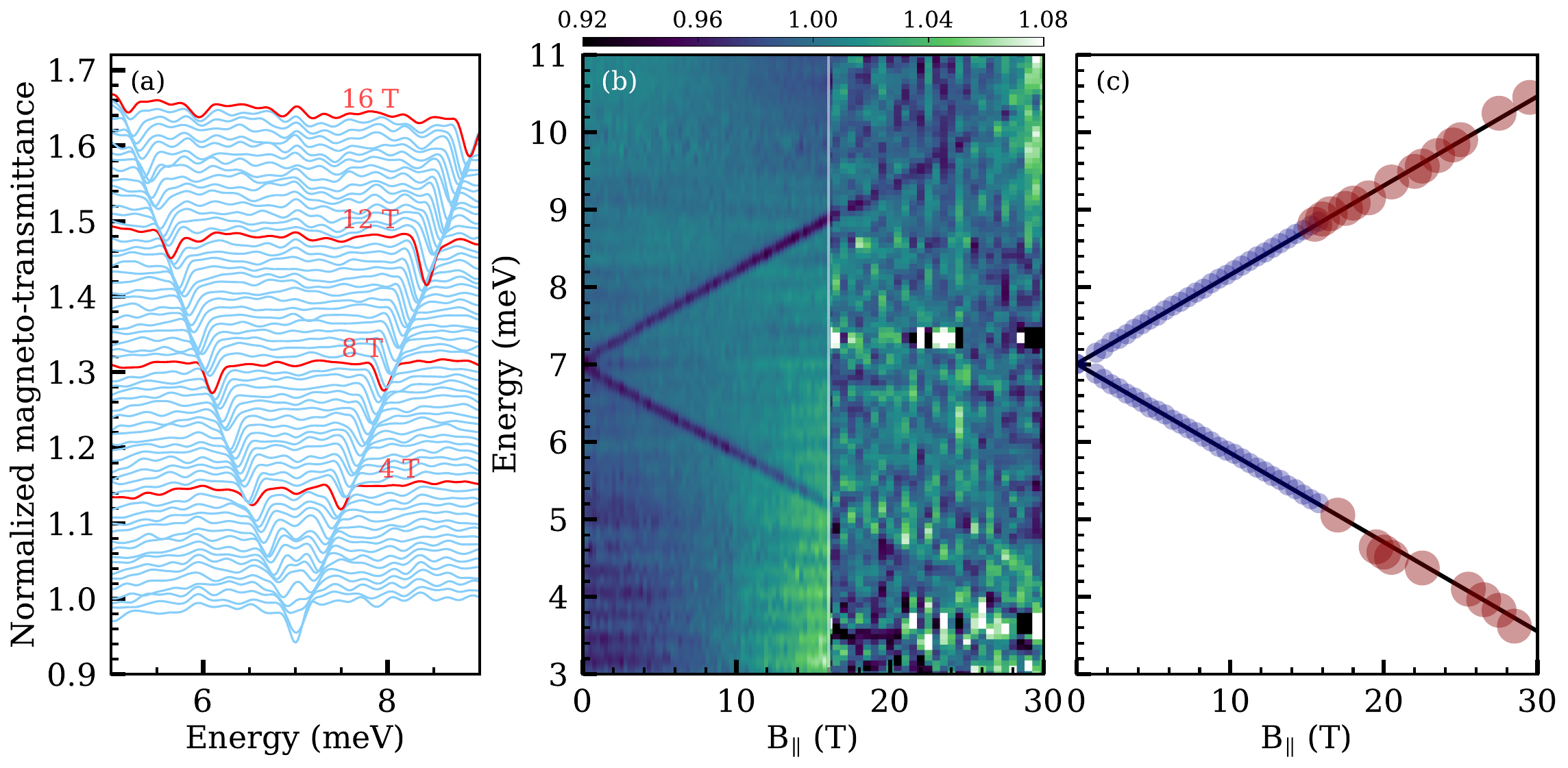}
    \caption{Infrared magneto-transmission of NaMnAs bulk crystal measured with the magnetic field applied along the tetragonal axis ($B_\|$) shown as a stack plot and false-color plot in (a) and (b), respectively. 
    The data were collected using a superconducting coil (up to \SI{16}{\tesla}) and a resistive coil (from \SI{16}{\tesla} to \SI{30}{\tesla}).
    Each spectrum is normalized by the average of spectra over the
    whole range of $B_\|$ scanned.  Panel (c) shows extracted positions of both AFMR branches, compared to theoretical expectations based on the Kittel's formula (\ref{eq:faraday_afmr}) for $g = 1.99$ (solid lines). The size of circles represents the error bar.}
    \label{fig:namnas_IR_faraday}
\end{figure*}

\section{Experimental results and discussion}
\label{Discussion}

Let us start the discussion of our experimental results with the THz-infrared transmission spectrum of NaMnAs measured $B = 0$ and $T = 4.2$~K,  with the light propagating along the $c$ axis of the material, see Fig.~\ref{fig:namnas_art_irphonons}a. Within the displayed spectral range, the transmission is modulated by a weak interference pattern, and the crystal remains largely transparent, except for two narrow, fully opaque regions, setting on at 15 and 25~meV. This agrees well with the expectations based on the tetragonal space group No.~129, determined based on X-ray scattering experiments~\cite{Achenbach1981-kl}. The symmetry analysis~\cite{RousseauJRS81,IvantchevJAC00,UmPRB12} suggests the presence
of two doubly degenerate $E_u$ phonon modes that are electric-dipole active in the given experimental configuration ($E$-field $\perp c$). When the wave vector is tilted away from the tetragonal axis, an additional out-of-plane $A_{2u}$ phonon line appears, see Fig.~\ref{fig:namnas_art_irphonons}. At higher frequencies, additional weaker excitations emerge, notably the one around 47~meV (denoted as $P$ in Fig.~\ref{fig:namnas_art_irphonons}). We attribute it to a multi-phonon resonance. 

\begin{figure*}[ht]
    \centering
    \includegraphics[width=.66\linewidth]{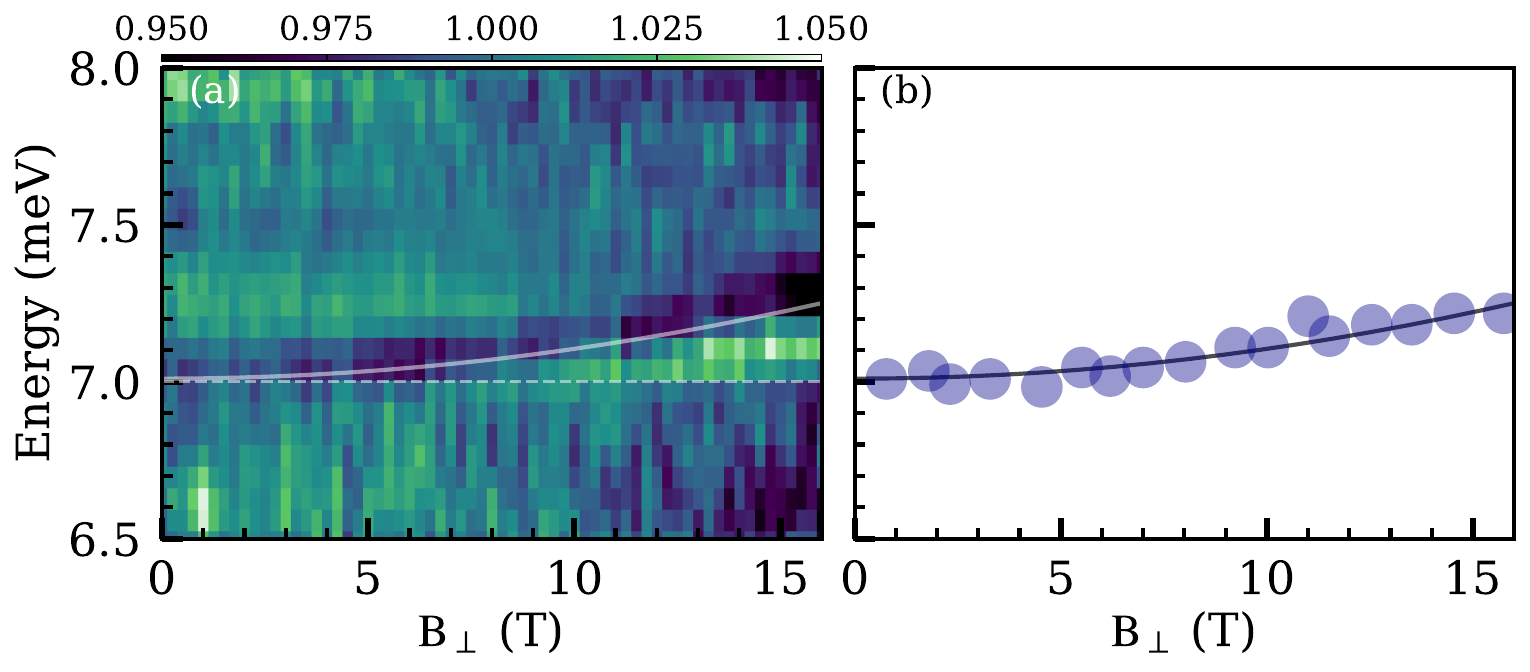}
    \caption{Panel (a): Color-plot of infrared magneto-transmission taken on a NaMnAs bulk crystal in the Voigt configuration,  with $B$ applied perpendicular to the tetragonal axis ($B_\perp$). Each spectrum is normalized by the average of spectra over the scanned magnetic field range (up to \SI{16}{\tesla}). The field-dependence from Eq. \eqref{eq:voigt_afmr} is shown as solid white line. The white dashed line shows the magnon energy at zero field and also the expected lower branch magnon energy.
    Panel (b): Extracted minima of the magneto-transmission along with the expected dependence of the magnon mode following Eq.~\eqref{eq:voigt_afmr} for $g=2$.}
    \label{fig:namnas_IR_voigt}
\end{figure*}

Magnetic-field experiments reveal another excitation active in the THz range in the response. In the limit of vanishing $B$, this excitation appears in the magneto-transmission spectrum at the energy of $\omega^0_{\mathrm{AFMR}}=(7.0\pm 0.2)$~meV, see
Figs.~\ref{fig:namnas_IR_faraday} and \ref{fig:namnas_IR_voigt}.

As discussed below in detail, the characteristic behavior of this excitation in the magnetic field applied along or perpendicular to the tetragonal axis allows us to assign it to AFMR -- in this case to a doubly degenerate magnon ($k = 0$) mode.  For $\mathbf{B} \| c$, the AFMR line splits symmetrically into two branches, both dispersing linearly in $B_\|$, see Fig.~\ref{fig:namnas_IR_faraday}. Such behavior is typical of AFMR in the easy-axis antiferromagnets. It is well described in the semiclassical framework by Kittel~\cite{KittelPR51,KefferPR52}, see also the review by Rezende et al.~\cite{Rezende2019}:
\begin{equation}
    \label{eq:faraday_afmr}
    \omega_{\mathrm{AFMR}}(B_\|) = \omega^0_{\mathrm{AFMR}}\pm g\mu_{B}B_\|.
\end{equation}
Comparing this formula with our data, see Fig.~\ref{fig:namnas_IR_faraday}c, we extracted the corresponding $g$ factor, \mbox{$g_\|=(1.99\pm0.05)$}, not far from the free electron value. Similarly to other easy-axis antiferromagnets~\cite{hagiwara_1999,VaidyaScience20}, a simple high-field extrapolation of the lower AFMR branch allows us to obtain a rough estimate of the spin-flop field of approximately 60~T.

For $\mathbf{B} \perp c$, the magnon modes' degeneracy is expected to  be removed again, but with a distinctively different behavior. When the magnetic anisotropy is relatively weak, which is often the case in manganese-based magnetic compounds~\cite{Szuszkiewicz2006-bu, Pepy1974-eb, hagiwara_1999, Wildes1998-rx} and which is \emph{a~posteriori} confirmed in our analysis, the upper mode is expected to increase monotonically with $B_\perp$:
\begin{equation}
    \label{eq:voigt_afmr}
    \omega_{\mathrm{AFMR}} (B_\perp) = \sqrt{
    \left(\omega^0_{\mathrm{AFMR}}\right)^2+(g \mu_\mathrm{B} B_\perp)^2},
\end{equation}
while the lower mode should remain nearly independent of the field: $\omega_{\mathrm{AFMR}} (B_\perp) \approx \omega^0_{\mathrm{AFMR}}$~\cite{KefferPR52}.

Experimentally, only a single branch is discerned in the data taken in the corresponding conformation, see Fig.~\ref{fig:namnas_IR_voigt}. It exhibits a monotonic blueshift with $B_\perp$, approximately quadratic in the low-field limit. The observed field-dependence is well reproduced using the above formula (\ref{eq:voigt_afmr}). The corresponding fit gives us $g$-factor of $(2.0 \pm 0.1)$. The lower branch does not clearly manifest in our data because  only the modes with non-negligible dispersion in magnetic field are visible in the relative magneto-transmission data. 

\begin{figure*}[ht]
    \centering
    \includegraphics[width=0.95\linewidth]{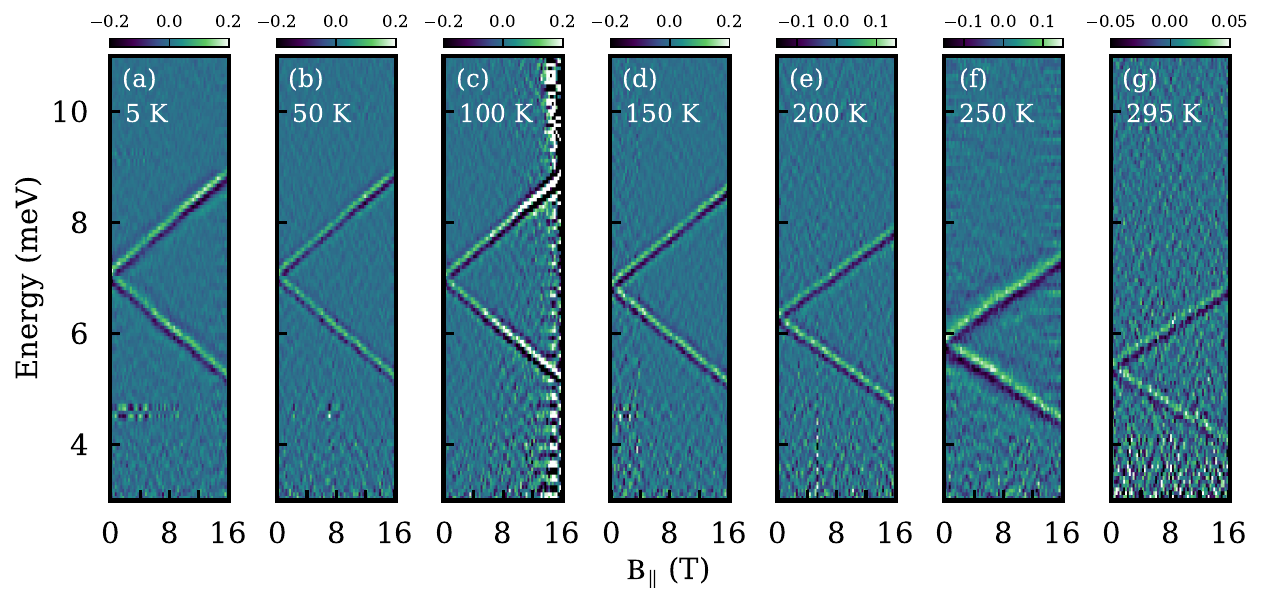}
    \caption{THz magneto-transmission of bulk NaMnAs, measured in the Faraday configuration at temperatures $T=5$, 50, 100, 150, 200, 250 and 295~K in panels (a-g), respectively. The magnetic field was applied along the tetragonal axis of the material. To plot the data, each individual spectrum collected at the magnetic field $B$ was normalized by the spectrum averaged over the interval $B\pm\delta B$ for $\delta B=5$~T, or shorter, respecting the range of magnetic fields explored ($0\leq B\leq16$~T). Then, the derivative of spectra with respect to the energy (frequency) was performed. With increasing $T$ the characteristic AFMR feature monotonically redshifts, from 7.0~meV at 5~K down to 5.4~meV at room temperature. The AFMR energies at $B=0$ were plotted as a function of $T$  in Fig.~\ref{fig:namnas_points_temp}, along with the corresponding effective $g$ factor, evaluated using Eq.~\eqref{eq:faraday_afmr}.}
    \label{fig:namnas_IR_temp}
\end{figure*}

Another set of experimental results, shown in Fig.~\ref{fig:namnas_IR_temp}, presents the temperature dependence of the magnon mode in the $\mathbf{B} \parallel c$ configuration. As the temperature increases from liquid-helium to room temperature, the characteristic pattern—consisting of two branches linear in $B$—remains unchanged. This indicates that the system preserves its easy-axis magnetic order throughout the entire temperature range investigated and confirms that the N\'eel temperature $T_\mathrm{N}$ of bulk NaMnAs exceeds room temperature.

A quantitative analysis reveals that the magnon mode softens monotonically with increasing temperature, decreasing from 7.0 to 5.4 meV as $T$ rises from 4.2 to 295 K at $B=0$. In a simple picture, this redshift reflects the reduction of the staggered magnetization with temperature, which ultimately vanishes at the ordering temperature $T_N$. In addition to the magnon energy, we observe a monotonic decrease of the $g$ factor with increasing $T$. Within the resolution of our spectra, no significant temperature-induced broadening of the AFMR line is detected. A more detailed discussion and modeling of the temperature dependence of both the magnon energy and the $g$ factor are presented in the following section.

It is worth noting that both magnon branches remain distant 
from phonon modes in the whole ranges of the magnetic field and temperature explored. Hence, we find no evidence of any magnon-phonon coupling, recently observed in several vdW antiferromagnets~\cite{LiuPRL21,VaclavkovaPRB21,ZhangCM21,CuiNC23}. The observed AFMR signal can thus be regarded as a textbook example of AFMR in easy-axis systems.

To describe the magnetic structure of NaMnAs let us invoke the effective spin Hamiltonian applied to NaMnAs in the preceding study by Volny et al.~\cite{VolnyPRB22}: 
\begin{equation}
\begin{split}
\label{eq:hamiltonian}
\hat{\mathcal H} = J_1 \sum_{{\langle ij\rangle}_1} {\bf S}_i\cdot {\bf S}_j +
J_2 \sum_{{\langle ij\rangle}_2} {\bf S}_i\cdot {\bf S}_j + J_3 \sum_{{\langle ij\rangle}_3} {\bf S}_i\cdot {\bf S}_j \\ + 
J_4 \sum_{{\langle ij\rangle}_4} {\bf S}_i\cdot {\bf S}_j
-D \sum_i (S_i^z)^2, 
\end{split}
\end{equation}
where the Heisenberg exchange interactions are included up to the fourth nearest neighbors -- both inter-sublattice $J_1$,$J_4$ and intra-sublattice $J_2$, $J_3$ couplings -- as well as the easy-axis-type single-ion anisotropy term.

The magnon dispersion corresponding to Hamiltonian~\eqref{eq:hamiltonian} can be calculated analytically using the standard linear spin wave theory, employing Holstein-Primakoff transformation. For a comparison with the AFMR experiment, only the $\Gamma$ point of the magnetic Brillouin zone is sufficient. Due to the coherent motion of spins within individual sublattices, the intra-sublattice couplings do not contribute to the magnon energy, reaching: 
\begin{equation}
    \label{eq:gamma}
    \omega_\Gamma = \omega^0_{\mathrm{AFMR}}= 2 S\sqrt{D(D+4J_1+8J_4)}.
\end{equation}



Knowing the N\'eel temperature of an antiferromagnet, the mean-field theory allows us to get a rough estimate of effective exchange coupling between sublattices, see, e.g., Ref.~\cite{MendiliCM26}:
\begin{equation}
    \label{eq:mean-field}
     k_B T_N \approx \frac{(S+1)S}{3}(4J_1-4J_2-2J_3+8J_4).
\end{equation}
Assuming that $J_1$ is the leading (antiferromagnetic) exchange interaction, as \emph{a posteriori} justified based on the results of DFT calculations, the N\'eel temperature of $T_N\approx 350$~K implies a rough estimate of $J_1^{\mathrm{exp}}\approx 4$~meV. The mean-field approximation, however, may overestimate the critical temperature by the factor up to or even exceeding 2, see Ref.~\cite{MendiliCM26}. This implies $J_1^{\mathrm{exp}}$ in the approximate 
range of 4-8~meV. Taking this, together with the experimentally determined AFMR energy, we obtain the estimate for the single-ion magnetic anisotropy, $D^{\mathrm{exp}} = 0.1-0.2$~meV. Nevertheless, since NaMnAs is a strongly anisotropic material -- with a large difference between in-plane ($J_1$ and $J_2$) and out-of-plane ($J_3$ and $J_4$) exchange coupling constants -- we take this result as a rough estimate only. 

Let us note that, in the estimate above, we considered the effective spin of $S=2$. This is lower as compared to expectations for a high-spin configuration of an isolated Mn$^{2+}$ ion ($S=5/2$), but in line with preceding studies of NaMnAs. The magnetic moment in NaMnAs, $\approx4\  \mu_B$, was reported by Bonger et al.~\cite{Bronger1986} and explained by the admixture of low and high spin states ($S=3/2$ and 1/2). This is due to the crystal field of Mn$^{2+}$ ions in NaMnAs that deviates from a pure tetrahedral one (of As atoms), see \cite{Achenbach1981-kl} for more details.

\section{Theoretical modelling}
\label{Theoretical_modelling}

Let us now confront our experimental results with expectations based on numerical simulations. We start with a model of phonon excitations, clearly visible in THz-infrared transmission (Fig.~\ref{fig:namnas_art_irphonons}). An ab initio simulation of the phonon band structure was performed with a focus on optically active ($k=0$) modes. These simulations were performed using the Quantum ESPRESSO package~\cite{GiannozziJPCM09} with LDA functionals and for different values of $U$. Norm-conserving pseudopotentials were taken from the "Pseudo Dojo" database. A kinetic energy cutoff of 80 Ry was used for the expansion of the wavefunctions. For each value of $U$, both the cell parameters and atomic positions were relaxed using a $13 \times 13 \times 8$ Monkhorst-Pack mesh, a force threshold of 2 Ry/Bohr, and a total energy convergence threshold of 2$\cdot10^{-10}$ Ry for ionic minimization. At each step, the self-consistent calculations were converged to a threshold of $10^{-10}$ Ry. Upon relaxation, infrared spectra were obtained from the effective charges, dielectric susceptibilities and phonons at the $\Gamma$ point using density functional perturbation theory (DFPT)~\cite{BaroniRMP01}. A convergence threshold of $10^{-16}$ Ry was used for the iterative solution of the DFPT equations. 

\begin{figure}[t]
    \centering
    \includegraphics[width=0.99\linewidth]{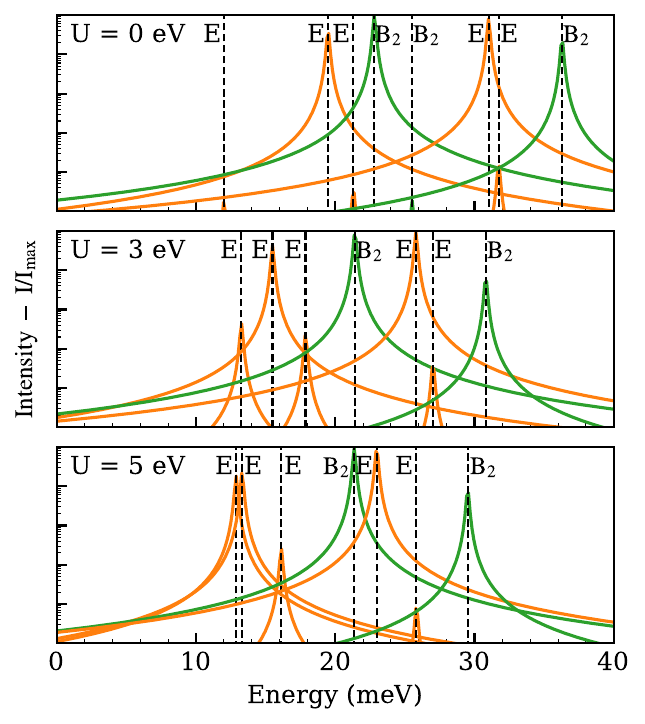}
\caption{Normalized IR spectra of NaMnAs based on DFT+U electronic structure (with $U=0,$ 3 and $5$ eV) computed using density functional perturbation theory. Peaks are labeled according to the irreducible representations of the point group $D_{2d}$ to which the corresponding infrared active phonon modes belong. The $E$ modes (orange) and $B_2$ (green) modes are active in the in-plane and out-of-plane components of the electric field, respectively.}
\label{fig-08}
\end{figure}

The results of these DFT calculations --- the intensity spectra that are proportional to the absorption coefficient --- 
are plotted in Fig.~\ref{fig-08} for three different values of $U=0,\ 3$ and 5~eV (as in Fig.~6 of Ref.~\cite{VolnyPRB22}). In order to capture all infrared active phonon modes, the calculations were made for the electric field of incoming radiation inclined by 45~degrees from the tetragonal axis of NaMnAs. Notably, the DFT calculations suggest a weak distortion of the crystal lattice. The tetragonal lattice remains preserved but the overall symmetry is reduced. In the terms of the point and space groups, the symmetry is reduced from $D_{4h}$ to $D_{2d}$ and No.~129 to 115, respectively. The corresponding Wyckoff positions are 2$g$ for Na as well as for As atoms. The Mn atoms are at 1$a$: (0,0,0) and 1$b$: (0.5,0.5,0). 

Comparing the numerically calculated phonon energies with the experimentally observed pattern in Fig.~\ref{fig:namnas_art_irphonons}, the best agreement is found for the largest considered value of $U=5$~eV. In a greater detail, the calculations predict: (i) a pair of strong, nearly degenerate in-plane $E$ modes around 13~meV; (ii) two strong and nearly degenerate modes (in-plane $E$ and out-of-plane $B_2$) around 22~meV and (iii) as compared to the previous ones, a relatively weaker out-of-plane mode $B_2$ at 30~meV. This matches relatively well with the pronounced phonon bands (denoted as $E_u$) observed in the transmission data at 15 and 25~meV, as well as with an additional weaker, out-of-plane active mode at 30~meV ($A_{2u}$). To avoid confusion, we recall that the notation in Fig.~\ref{fig:namnas_art_irphonons} ($E_u$ and $A_{2u}$) follows the space group No.~129, determined experimentally for NaMnAs by X-ray scattering~\cite{Achenbach1981-kl} while the calculations predict a lower symmetry, No.~115, with optically active phonon modes denoted as $E$ and $B_2$.



Setting approximately the value of $U\approx5$~eV, we took the corresponding parameters of the relaxed unit cell and calculated the exchange coupling in NaMnAs. These additional \emph{ab initio} calculations were performed with the OpenMX package. The simulations were carried out using local spin density functional of Ceperley-Alder. The exchange interactions were computed in TB2J extension of OpenMX. This code uses the magnetic force theorem, taking the local rigid spin rotations as a perturbation. The calculation uses the Green's function method, utilizing the DFT Hamiltonian in the local atomic orbitals basis set used by OpenMX. The output of the calculation had to be renormalized to $1/S^2$ to be in accordance with our Hamiltonian \eqref{eq:hamiltonian} ($S=2$).


\renewcommand{\arraystretch}{1.2}
\begin{table}[ht]
    \caption{Theoretically calculated microscopic parameters (DFT calculation using the OpenMX software and TB2J method) for NaMnAs in relaxed unit cell calculated for $U=5$~eV.}
     \vspace{2mm}
    \centering
    \begin{tabular}{|l|cl|}
    \hline
    J\textsubscript{1} & &7.55~meV\\
    J\textsubscript{2} & &1.84~meV \\
    J\textsubscript{3} & &-0.03~meV \\
    J\textsubscript{4} & &0.07~meV \\ 
    \hline
    \end{tabular}
    \label{tab:EE}
\end{table}

\begin{figure*}[ht]
    \centering
    \includegraphics[width=0.75\linewidth]{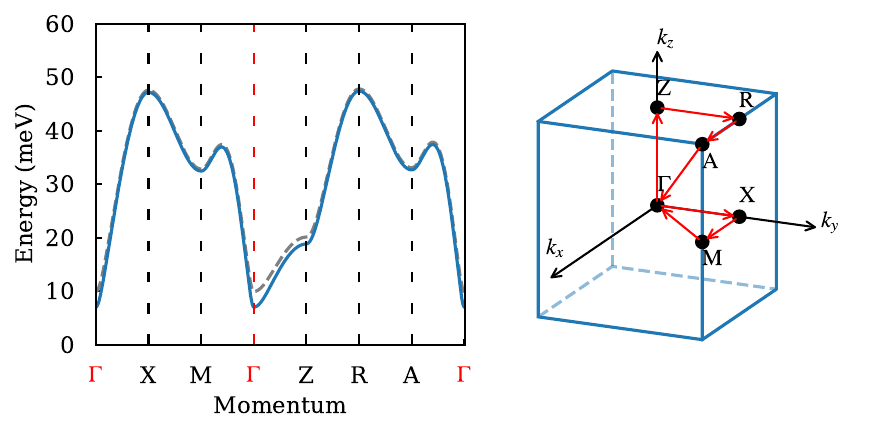}
    \caption{Left panel: magnon dispersion in NaMnAs for $B=0$. The dispersions are plotted for the parameters determined by ab initio calculations for $U=5$~eV (Tab.~\ref{tab:EE}) and two limiting values of the single-ion magnetic anisotropy $D=0.1$ and 0.2~meV -- the range estimated from our experimental data in the main text. The dispersion was calculated using SpinW package for MATLAB for the given set of microscopic parameters. Right panel: the BZ for a tetragonal crystal lattice with relevant high-symmetry points highlighted. The red arrows represent the $k$-cut directions used in the magnon dispersion.}
    \label{fig:namnas-dispersion}
\end{figure*}

The outcome of these calculations, i.e., the theoretical estimates of $J_i$ (cf. Fig.~\ref{fig:namnas-structure}) are listed in Tab.~\ref{tab:EE}, with the precision corresponding to the accuracy of the numerical method. The calculated strength of $J_1$ coupling falls well in the range 4-8 meV estimated from $T_N$ using the mean-field theory, see Sec.~\ref{Discussion}. The DFT+U results are also in-line with our previous assumption that $J_1$ is the leading exchange coupling parameter. Motivated by this reasonable agreement between numerical calculations and our experimental findings, we plot (Fig.~\ref{fig:namnas-dispersion}) the theoretically expected magnon dispersions using the calculated $J_i$ for two limiting values of the magnetic anisotropy $D=0.1$ and 0.2~meV. To this end, the SpinW package for MATLAB~\cite{TothJPCM15} was used. As expected the magnon dispersion only changes with the varying magnetic anisotropy in the vicinity of the $\Gamma$ point. This theoretically proposed magnon dispersion calls for a comparison with future inelastic neutron scattering experiments.



\begin{figure}[t]
    \centering
    \includegraphics[width=1.\linewidth]{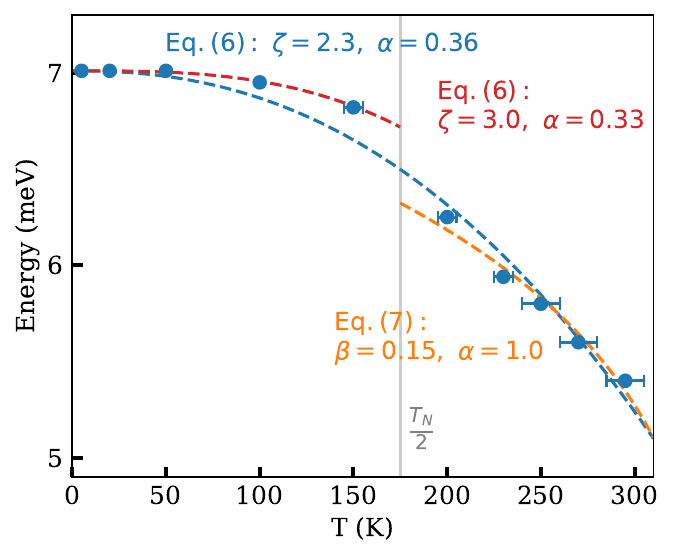}
    \caption{The magnon energy $\omega_{\Gamma}$ 
    measured at selected values of temperature in the range 4.2-295~K. The dashed lines are described in detail in Sec.~\ref{Theoretical_modelling}. The dashed blue line shows the fit using \eqref{eq:beta-alpha} for the whole temperature range. The dashed red line shows the fit using \eqref{eq:beta-alpha} in the low-temperature regime, dominated by quantum fluctuations. The dashed orange line shows the fit using \eqref{eq:critical_exp} at high temperatures, in the regime driven by thermal fluctuations.}
    \label{fig:namnas_temp-exp}
\end{figure}

Let us now turn our attention at AFMR experiments at elevated temperatures, see Figs.~\ref{fig:namnas_IR_temp} and \ref{fig:namnas_temp-exp}. These show a monotonic decrease of the magnon energy with increasing $T$. Taking the AFMR energy as a measure of the staggered magnetization, we started with comparing the experimental AFMR energies with the simple analytical formula~\cite{Kobler2020-sz}:
\begin{equation}
    \frac{\omega_{\mathrm{AFMR}}(T)}{\omega_{\mathrm{AFMR}}(T=0)}= \left[1-\alpha\left(\frac{T}{T_N}\right)^\zeta\right],
    \label{eq:beta-alpha}
\end{equation}
This formula comprises the scaling factor $\alpha$ and the parameter $\zeta$ that can be interpreted as a critical exponent. The N\'eel temperature was in this case fixed at $T=350$~K. The best fit, see blue dashed line in Fig.~\ref{fig:namnas_temp-exp}, yields the parameters $\zeta=2.3$ and $\alpha=0.36$. 
Consistent with previous studies ~\cite{Kobler2020-sz}, we find that the full temperature range investigated cannot be adequately described using a single power law formula with parameters $\zeta$ and $\alpha$.  

Therefore, we decided to consider two regimes separately, below and above the temperature of $T_N/2$ around which the $T$-dependence of the magnon energy seems to change its character rather abruptly, see Fig.~\ref{fig:namnas_temp-exp}. The power laws used to reproduce the data in these two regimes then differ. At low temperatures, one may expect that those are quantum fluctuations that dominate the temperature dependence. This results in a relatively weak redshift of the magnon energy, i.e., $\zeta >1$ in ~Eq.~\eqref{eq:beta-alpha}. In contrast, as $T$ approaches $T_N$, thermal fluctuations should become significant, leading to an accelerated temperature-driven redshift of the magnon energy ($\beta <1$): 
\begin{equation}
    \frac{\omega_{\mathrm{AFMR}}(T)}{\omega_{\mathrm{AFMR}}(T=0)}=\alpha \left[1-\frac{T}{T_N}\right]^\beta.
    \label{eq:critical_exp}
\end{equation}
Accordingly, we obtain $\zeta = 3.0$, $\alpha = 0.33$ in the low-temperature regime and $\beta = 0.15$, $\alpha = 1.0$ at higher temperatures (see the red and orange dashed lines in Fig.~\ref{fig:namnas_temp-exp}, respectively). These coefficients are consistent with the already-known properties of NaMnAs that is an anisotropic, quasi-2D antiferromagnet~\cite{Kobler2005-to, Kobler2010Book}. Let us note, however, that the curves fitted at low- and high-temperature regimes do not cross each other. This may suggest the existence of an additional, intermediate regime of the boson field~\cite{Kobler2005-to,Kobler2010Book,Kobler2020-sz} at temperatures 150\nobreakdash-200~K. 

Although the above empirical formula compares reasonably well with the experimental temperature dependence of the magnon energy, it does not account for the temperature dependence of the $g$ factor. To reproduce it, we have turned to a numerical solution. To this end, the Heisenberg exchange interactions in Eq.~\eqref{eq:hamiltonian} have been approximated by temperature dependent local magnetic fields, one for each spin. In a greater detail, the local mean-field Hamiltonians read: $h_i= -g\mu_B\vec{b}_i(T)\cdot\vec{S}_i-D (S^z_i)^2$, where the local magnetic fields $\vec{b}_i(T)$ fulfill the self-consistent equations: 
$g\mu_B\vec{b_i}(T)=g\mu_B\vec{B}-\sum_{j\ne i} J_{ij}\langle\vec{S}_j\rangle$, with the temperature-dependent local magnetizations $\langle\vec{S}_i\rangle$ computed using the local Hamiltonians $h_i$. This mean-field approach allows us not only to compute the thermal ground state for every value of $B$ and $T$, but also, when the equations of motion of the spin operators $\vec{S}_i$ are linearized, the temperature dependence of the magnon dispersion. More precisely, we obtain the two magnon branches, like in Fig.~\ref{fig:namnas_IR_temp}, $\omega_{\pm}(T,B)$ as function of both $T$ and $B$. Thereby, the value at $B=0$, provides the standard linear spin wave theory solution $\omega_\Gamma(T)$ and the effective $g_{\mathrm{eff}}(T)$ is obtained from a linear regression (as a function of $B$) of the two magnon energies. 

\begin{figure}
    \centering
    \includegraphics[width=1.\linewidth]{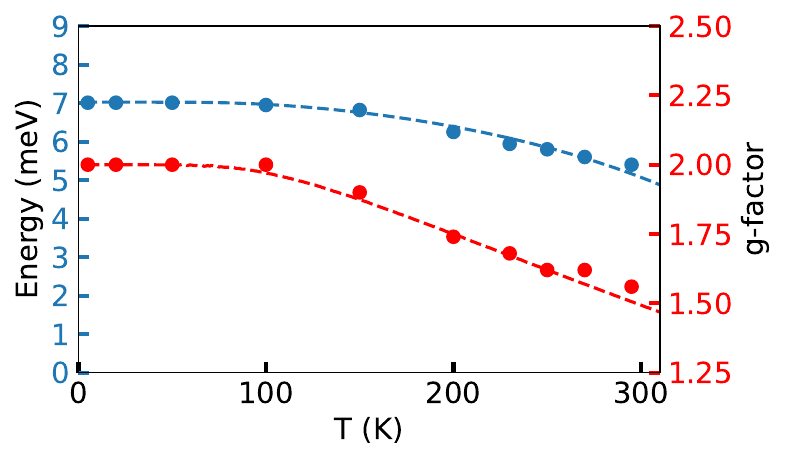}
    \caption{Magnon energy, $\omega_{\Gamma}$, (blue) and effective $g$ factor (red) as functions of the temperature. The filled circles are the experimental data points taken from Fig.~\ref{fig:namnas_IR_temp}, whereas the dashed lines are the mean-field computation with a proper rescaling in the temperature dependence of magnon energy. In both cases the agreement is quite good emphasizing that the effective model~\eqref{eq:hamiltonian} is a fair description of the magnetic properties of NaMnAs.}
    \label{fig:namnas_points_temp}
\end{figure}

It is well known, however, that the mean-field approaches, do not properly capture some of thermal fluctuations, resulting in incorrect values of the N\'eel temperature. A simple way to take this difficulty into account is to rescale linearly the temperature axis of the calculated mean-field solution to better fit them to the experimental data. The comparison between the experimental data and the rescaled theoretical curves is displayed in Fig~\ref{fig:namnas_points_temp}. Good agreement is found not only for $\omega_{\Gamma}$, but also for the $g$-factor.
This emphasizes that the Hamiltonian~\eqref{eq:hamiltonian}, within the mean-field approach, fairly well captures the magnetic properties of NaMnAs. 

Nevertheless, one has to note that this agreement has been obtained using the rescaling 
coefficient of 3/4 for the temperature axis of $\omega_{\Gamma}$. No rescaling was applied to the $g$ factor dependence.  
On the one hand, the magnon energies roughly correspond to the singularities/maxima of the transverse susceptibility $\chi_{\perp}(\omega,T)$. On the other hand, the $g$ factor probes how $\chi_{\perp}(\omega,T)$ itself changes as a function of the magnetic field $B$. Therefore, one may expect that $\omega_{\Gamma}$ and the $g$ factor are impacted differently by the thermal fluctuations, leading to different deviations as functions of the temperature with respect to the mean-field computations. There are a few ways of improving the mean-field, in particular accounting better for the thermal fluctuations due to the magnons and their interactions, but they lead to complex numerical computations and are, therefore, beyond the scope of this work~\cite{RajeevPavizhakumari_2025}.

\section{Conclusions}

In summary, we have studied antiferromagnetic resonance in tetragonal NaMnAs in a broad range of temperatures and applied magnetic fields. According to theoretical expectations (DFT) and in line with experiments performed on this material so far, the observed response is consistent with NaMnAs being an easy-axis antiferromagnet with $T_N$ above the room temperature. The energy of the observed $k=0$ mode falls into the THz spectral range. Hence, NaMnAs qualifies as another interesting material for THz-based applications and research, operating at ambient temperature. Moreover, this material is exfoliable, possibly down to the monolayer limit. The extracted single-ion magnetic anisotropy is relatively large. It exceeds the anisotropies typically found in manganese-based antiferromagnets such as MnTe~\cite{Szuszkiewicz2006-bu,DzianPRB25}, MnO~\cite{Pepy1974-eb}, MnF$_2$~\cite{hagiwara_1999} or MnPS$_3$~\cite{Wildes1998-rx}, but it is smaller in comparison with rare-earth manganites \cite{HolmPRB18}.

\section*{Acknowledgement}
Authors appreciate the discussion with M. Zhitomirsky. J.D. and M.O. acknowledge the support received through the ANR-22-EXES-0001 project. Discussions with Alberto Marmodoro are gratefully acknowledged. S.R. and R.M. acknowledge the CINECA award under the ISCRA initiative, for the availability of high performance computing resources and support. We acknowledge support from the Dioscuri Program LV23025 funded by MPG and MEYS, GACR grant 25-18244S. We also acknowledge the support from OPJAK–Ferroic multifunctionalities (FerrMion CZ.02.01.01/00/22\_008/0004591). This work was supported by the Ministry of Education, Youth and Sports of the Czech Republic through the e-INFRA CZ (ID:90254). This work was co-funded by the European Union and the Czech
Ministry of Education, Youth and Sports (Project TERAFIT -
CZ.02.01.01/00/22\_008/0004594). This work was supported by the Charles University grant SVV–2025–260836.
The data that support the findings of this article are openly available \cite{Veis2026-ns}.



\bibliography{namnas}

\end{document}